\documentclass[aps,prc,twocolumn,amsmath,superscriptaddress,floatfix,nofootinbib]{revtex4-2}

\usepackage{float} 
\usepackage{setspace,ulem, epsfig,amssymb,amsfonts,amsmath,mathtools,bm,color,xcolor,graphicx,braket,adjustbox,esint,upgreek,verbatim,ctable}

\usepackage[colorlinks, linkcolor=red,anchorcolor=green,citecolor=blue]{hyperref}

\newcommand{\tred}{\textcolor{black}}

\begin{document}

%\title{Heavy flavor collective flow in small collision systems}
\title{Exploring parameter dependence of heavy-flavor dynamics in small collision systems}
\author{Grace Pang}\email{gpang@ucdavis.edu}
\affiliation{Department of Physics and Astronomy, University of California, Davis, Davis, California 95616, USA}

\date{\today}
\begin{abstract}
\tred{Observations from high-multiplicity proton-lead ($p$-Pb) collisions indicate that small systems may exhibit collective behavior in both heavy and light hadrons. This work investigates the roles of initial- and final-state interactions in shaping the nuclear modification factor and elliptic flow of $D$ mesons measured in $p$-Pb collisions. Initial-state effects, including the Cronin and shadowing effects, are considered in the heavy-quark initial conditions, while final-state interactions are simulated through Langevin evolution combined with the coalescence model of hadronization. Different initial geometries attributed to fluctuations in the medium's energy density are parametrized and translated into the momentum anisotropies of both light and heavy quarks. The corresponding $R_{pPb}$ and $v_2$ of D mesons in 8.16 TeV $p$-Pb collisions are calculated under different assumptions for the final-state interactions. Assuming that the initial-state effects only modify the transverse momentum spectra without altering the azimuthal distribution of heavy quarks, the measured $R_{pPb}$ of D mesons can be qualitatively reproduced by the combined influence of initial- and final-state effects. However, the observed $v_2$ cannot be accounted for by final-state interactions alone. These results suggest that additional contributions to azimuthal anisotropies of heavy quarks originating from initial-state effects are required to explain the experimentally observed $v_2$. 
}

\end{abstract}
\maketitle

\section{Introduction} 
In heavy-ion collisions at the Relativistic Heavy Ion Collider (RHIC) and the Large Hadron Collider (LHC), a deconfined state of strongly interacting matter, the Quark-Gluon Plasma (QGP), is believed to be formed, governed by quantum chromodynamics (QCD)~\cite{Bazavov:2011nk}. Extensive studies have been conducted to investigate the phenomena exhibited by this deconfined matter~\cite{Heinz:2013th,He:2022ywp,Zhao:2020jqu,Zhou:2023pti}.

A number of constituent quark (NCQ) scaling behavior has been observed in the collective flow of light hadrons~\cite{PHENIX:2006dpn,ALICE:2014wao}. This suggests that the degrees of freedom in the hot medium are primarily characterized by partons during their collective expansion~\cite{Dunlop:2011cf}. In both proton–nucleus (pA) and nucleus–nucleus (AA) collisions, the spatial anisotropy of the initial energy density is transformed into momentum anisotropy in a phemomenon well-described by hydrodynamic equations~\cite{Weller:2017tsr}, where the linear and cubic response to the initial eccentricity of the medium determines the anisotropy of the light hadron momentum distribution ~\cite{Gardim:2011xv,Noronha-Hostler:2015dbi}.

Due to the dependence of light hadron final azimuthal anisotropy on the spatial anisotropy of the initial energy density, considering fluctuations in initial energy density is critical for understanding the collective flow of light hadrons in heavy-ion collisions~\cite{Socolowski:2004hw}. Event-by-event fluctuations have been studied to explain the puzzle of the nuclear modification factor $R_{AA}$ and elliptic flow $v_2$ in jets~\cite{Noronha-Hostler:2016eow}. Meanwhile, triangular flow $v_3$, originating only from initial fluctuations, has been observed in both light hadrons~\cite{Alver:2010gr} and heavy hadrons~\cite{Zhao:2021voa}. Initial geometry of the medium with such fluctuations can be simulated with multiple ``hot spots'' with estimated typical sizes of around 0.2 fm~\cite{Zhao:2022ugy}.

 In small collision systems at both LHC and RHIC energies, an increasing number of experimental and theoretical studies suggest that collective flows are also formed in p+A, d+Au, and even $\gamma^*+A$ collisions~\cite{Zhao:2022ayk,Zhao:2022ugy,ATLAS:2021jhn,PHENIX:2018lia,STAR:2022pfn}. In particular, an approximate NCQ scaling of light hadron $v_2$ at intermediate transverse momenta $p_T$ is observed in $p$-Pb collisions, indicating partonic degrees of freedom within the medium. This behavior can be described by the quark coalescence model of hadronization~\cite{Zhao:2020wcd}.

On the subject of hard probes, the puzzle of jet $R_{AA}$ and $v_2$ has been studied with the contributions of jet energy loss, quark coalescence, and hadron cascade~\cite{Zhao:2021vmu}. This sheds light on the understanding of $D$ meson $R_{pPb}$ and $v_2$ in high multiplicity $p$-Pb collisions at 8.16 TeV and 5.02 TeV, respectively~\cite{CMS:2018loe,ALICE:2019fhe}. 
The effects of the hot deconfined medium on heavy quarks and quarkonium have been observed in \( p \)-Pb collisions. The ground and excited states of charmonium~\cite{ALICE:2015kgk,Chen:2016dke} and bottomonium~\cite{CMS:2022wfi,Chen:2023toz} exhibit different levels of suppression, which is attributed to distinct final-state interactions and cannot be explained solely by cold nuclear matter effects. 

Within the deconfined medium produced in \( p \)-Pb collisions, heavy quarks undergo significant energy loss due to strong coupling with the bulk medium~\cite{Xing:2024qcr}. 
\tred{Various approaches have been developed to address the significant elliptic flow observed for heavy flavor particles in these collisions. The AMPT model attributes both \( R_{pPb} \) and \( v_2 \) of \( D \) mesons to the combination of a strong Cronin effect and final-state interactions~\cite{Zhang:2024zga}. In this framework, random scatterings between heavy quarks and thermal partons enhance the elliptic flow of \( D \) mesons, but suppress their \( R_{pPb} \) in the intermediate \( p_T \) region. Inclusion of the Cronin effect, which increases the initial transverse momentum of heavy quarks, improves agreement with experimental \( D \) meson \( R_{pPb} \) data. 
Alternatively, in the Color Glass Condensate model, the significant \( v_2 \) of \( J/\psi \) and \( D \) mesons is attributed to initial state effects, which generate azimuthal angular correlations between heavy quarkonia and charged light hadrons in proton-nucleus collisions~\cite{Zhang:2019dth,Zhang:2020ayy}. In Ref.~\cite{Soudi:2023epi}, it is proposed that the initial transverse momentum distributions of unpolarized and polarized partons inside unpolarized nucleons can lead to the observed azimuthal anisotropy of high \( p_T \) hadrons in $p$-Pb collisions, without modifying the angle-integrated spectra. 
Together, these studies suggest different possible origins of the observed \( v_2 \) of heavy flavor particles in small collision systems, highlighting the need for a detailed investigation of the parameter dependence of heavy-flavor elliptic flow in $p$-Pb collisions.
}

\tred{This work investigates the impacts of various factors on the \( v_2 \) and \( R_{pPb} \) of \( D \) mesons in $p$-Pb collisions, with a particular focus on cold nuclear matter effects (Cronin and shadowing), and final-state interactions within the anisotropic medium produced in $p$-Pb collisions. The initial spatial anisotropy of the medium is translated into the azimuthal anisotropy of light-parton momentum distributions, which can be transferred to heavy quarks through random scatterings. \( D \) mesons formed via coalescence of heavy quarks with thermal light quarks can therefore inherit elliptic flow through these final-state interactions. 
Additionally, heavy quarks may acquire momentum anisotropy due to the path-length-dependent energy loss experienced by quarks propagating along different trajectories within the anisotropic medium. By incorporating both cold nuclear matter effects and small QGP droplets with different eccentricities, the \( v_2 \) of \( D \) mesons is computed using the Langevin approach in combination with the coalescence model.
}

This paper is organized as follows. Section II introduces the Langevin equation and the coalescence model used to study the dynamical evolution of heavy quarks in the deconfined medium and the subsequent hadronization process. The hot medium is generated using hydrodynamic simulations with different initial conditions, described in Section III. In Section IV, the nuclear modification factors and elliptic flows of $D$ mesons within different fluctuating media are presented. A summary is provided in Section V.

\section{theoretical model}

\subsection{Dynamical evolutions in the medium}
Heavy quarks are strongly coupled to the small deconfined medium produced in $p$-Pb collisions. Due to their large masses relative to the medium temperature, heavy quark production and annihilation through thermal processes are suppressed, allowing their trajectories to be tracked throughout the medium's evolution.

Transport equations~\cite{Zhang:2025cvk} and the Fokker-Planck equation~\cite{Walton:1999dy,He:2013zua} have been developed to study the evolution of the phase space distribution of heavy quarks. Assuming small momentum transfer between heavy quarks and thermal light partons during each interaction, the dynamical evolution of heavy quarks can be effectively described by the classical Langevin equation~\cite{He:2014cla,Cao:2018ews,Chen:2021akx,Yang:2023rgb}:
 \begin{align}
\label{lan-gluon}
{\frac{d{\bf p}_Q}{dt}}= -\eta(p_Q) {\bf p}_Q +{\bf \xi} + {\bf f}_g.
\end{align}
where \( {\bf p}_Q \) is the heavy quark momentum, \( \eta \) is the drag coefficient, the random noise term \( \xi \) represents random kicks from the thermal medium, and ${\bf f}_g$ accounts for force from medium-induced gluon radiation. The drag term is given by \( \eta = {\kappa}/{(2TE_Q)} \), which depends on the heavy quark energy \( E_Q = \sqrt{m_Q^2 + p_Q^2} \) and the medium temperature \( T \).  The momentum diffusion coefficient is \( \kappa = {2T^2}/{\mathcal{D}_s} \). Recent studies and results from lattice QCD calculations suggest a small value for the spatial diffusion coefficient \( \mathcal{D}_s \), especially in temperature regions near the critical deconfined phase transition temperature \( T_c \). In this work, \( \mathcal{D}_s(2\pi T) = 3 \) is used, representing a strongly coupled medium~\cite{Banerjee:2011ra} according to deep learning ~\cite{Guo:2023phd} and  Bayesian analyses~\cite{Xu:2017obm}. This value is slightly smaller than those commonly employed to study large collision systems~\cite{Li:2024wqq}.

Previous studies have neglected the momentum dependence of the white noise term,  \( {\bf \xi} \), which is determined via the relation:
\begin{align}
\langle \xi^{i}(t)\xi^{j}(t^\prime)\rangle =\kappa \delta ^{ij}\delta(t-t^\prime), 
\end{align}
with $i,j=(1,2,3)$ being the index of the three spatial dimensions. The medium-induced parton radiation also introduces a force \( {\bf f}_g = - \frac{d{\bf p}_g}{dt} \), where \( {\bf p}_g \) is the momentum of the emitted gluon. \tred{In the numerical calculations, when the time step \( \Delta t \) is sufficiently small, the probability of emitting a single gluon in this time step is assumed to be~\cite{Cao:2013ita}: \(
P(t, \Delta t) = \Delta t \int dx \, dk_T^2 \, \frac{dN_g}{dx \, dk_T^2 \, dt}.\)
Here, \( x = E_g / E_c \) is the ratio of the gluon energy to the charm-quark energy, and \( k_T \) is the transverse momentum of the emitted gluon. The spectrum of the emitted gluon, \( \frac{dN_g}{dx \, dk_T^2 \, dt} \), is calculated using perturbative QCD~\cite{Zhang:2003wk, Majumder:2009ge}.}

As the system cools, heavy quarks undergo a hadronization process when they exit the deconfined medium. The coalescence model has been developed to model this dynamical process phenomenologically, and has successfully described the observed experimental spectra of light hadrons in previous studies~\cite{Zhao:2020wcd}. The \( D \) mesons produced from this process consist of one heavy quark and a light anti-quark, with a momentum distribution given by~\cite{Greco:2003vf,Chen:2021uar}:
\begin{align}
    \frac{dN}{d^3{\bf P}_M}=&g_M\int d^3{\bf x}_1 d^3{\bf p}_1d^3{\bf x}_2d^3{\bf p}_2 f_Q({\bf x}_1, {\bf p}_1) f_{\bar q}({\bf x}_2, {\bf p}_2) \nonumber \\
    &\times W_M({\bf x}_r, {\bf p}_r)\delta^3 ({\bf P}_M-{\bf p}_1-{\bf p}_2),
    \label{lab-coal}
\end{align}
where \( g_M \), the statistical factor of the meson, is replaced with the hadronization ratio, \( g_M \equiv H_{c \rightarrow D^0} \) representing the fraction of charm quarks that transition into a specific state of the \( D \) meson. These values are taken to be \( H_{c \rightarrow D^0} = 9.5\% \) and \( H_{c \rightarrow D^{*0}} = 20\% \)~\cite{ALICE:2018lyv}. \( f_Q \) and \( f_{\bar{q}} \) are the momentum distributions of charm and light quarks, respectively. \( f_Q \) is obtained through the Langevin equation, and \( f_{\bar{q}} \) is assumed to follow a Fermi distribution with a fixed temperature \( T_c = 0.16 \) MeV and a light quark mass of 0.3 GeV. \( W_M({\bf x}_r, {\bf p}_r) \) is the normalized Wigner function of the \( D^0 \) meson, modeled as a Gaussian function with Gaussian width $\sigma_M^2$, which is related to the mean-square radius \( {\langle r^2 \rangle_D} \) of the meson by: $\sigma_M^2={\frac{4}{3}} {\frac{(m_1+m_2)^2}{m_1^2+m_2^2}}\langle r^2\rangle_D$. For both \( D^0 \) and \( D^{*0} \) mesons, \( {\langle r^2 \rangle_D} = 0.43^2\rm\ {\rm fm^2} \).
The relative position and momentum of the pair of quarks in their center-of-mass frame, \( {\bf x}_r \) and \( {\bf p}_r \), are defined as:
\begin{align}
    {\bf x}_r &= x_1^{\rm cm}-{\bf x}_2^{\rm cm}\nonumber \\
    {\bf p}_r & = \frac{E_2^{\rm cm}{\bf p}_1^{\rm cm}- E_1^{\rm cm}{\bf p}_2^{\rm cm}}{E_1^{\rm cm} + E_2^{\rm cm}}.
\end{align}
$E_1^{\rm cm}$ and $E_2^{\rm cm}$ are the center-of-mass energies of the heavy and light quarks, respectively. Using the Langevin-evolved final momentum distributions of charm quarks, the resulting \( D\) meson spectrum is computed with this coalescence model.

\subsection{Spatial and momentum anisotropy}
In relativistic heavy-ion collisions, it has been established that the final momentum anisotropy of the hadrons reflects the initial spatial anisotropy of the QGP medium. 
The initial spatial anisotropy in energy density is transformed into hadron final momentum anisotropy, with the transform efficiency determined by the transport coefficients of the hot medium~\cite{Teaney:2010vd}. For light hadrons, a significant elliptic flow $v_2$ ~\cite{Zhao:2017yhj} is attributed to the geometric shape of the overlap between two nuclei. In small collision systems, the evident elliptic flow of light hadrons has been observed and well described with the hydrodynamic model~\cite{Zhao:2020wcd}. The spatial anisotropy distributions of energy density introduced will be used to examine the effects of different initial anisotropies on final momentum anisotropy of heavy quarks and resulting D mesons after moving out of the small QGP droplet in \( p \)-Pb collisions. ~\cite{CMS:2018loe}.

The final-state azimuthal momentum anisotropy of the $D$ mesons is quantified in terms of the coefficients $v_n$ of the Fourier expansion of the particle azimuthal distribution~\cite{Heinz:2013bua},
\begin{align}
    \frac{dN}{d\phi}&\propto 1+2\sum_n v_n \cos[n(\phi-\Psi_n)]\\
    v_n &=\langle\langle \cos2(\phi-\Psi_2)\rangle\rangle=\frac{\int d\phi\cos[n(\phi-\Psi_n)]\frac{dN}{d\phi}}{\int d\phi \frac{dN}{d\phi}}
\end{align}
where the event-plane angle $\Psi_n$ can be determined from the distribution of light hadrons by $\Psi_n=(1/n)\arctan(\langle p_T\sin n\phi\rangle/\langle p_T\cos n\phi\rangle)$~\cite{Niemi:2012aj}.

The elliptic flow $v_2$ is is known to correlate with the initial spatial eccentricity $\epsilon_2$ in the transverse plane ${\bf x}_T=(x,y)$~\cite{Drescher:2006pi,Bhalerao:2011yg}, which is defined as~\cite{Bhalerao:2011yg},
\begin{align}
    \epsilon_n \equiv  \frac{\int r^n \cos[n(\phi-\Psi_n)] e(r, \phi) rdr d\phi}{\int r^n e(r, \phi) rdr d\phi}.
\end{align}
Here,  $e(r,\phi)$ is the initial energy density. In this study, different anisotropies characterized with varying values of eccentricity $\epsilon_2$ (with n=2 in the above equation) are used to calculate the corresponding $D$ meson final flow.

\subsection{Initial conditions}
\begin{figure}[!hbt]
    \centering
\includegraphics[width=0.42\textwidth]{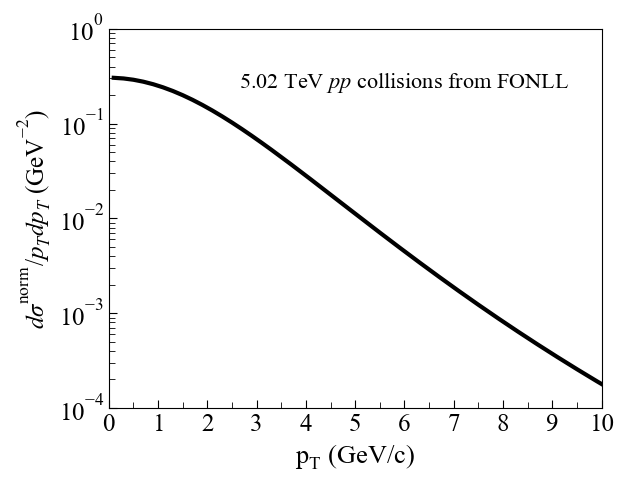}
    \caption{  The \tred{normalized} initial transverse momentum distribution of charm quarks in the central rapidity bin of 5.02 TeV $pp$ collisions from the FONLL model. }
    \label{lab-fig-dnptdpt}
\end{figure}

To solve the Langevin equation for heavy-quark evolution in \( p \)-Pb collisions, the initial transverse-momentum distribution of the quarks is required. \tred{This distribution of charm quarks is obtained from the "Fixed Order + Next-to-Leading Log" (FONLL) framework.~\cite{Cacciari:2001td}. To incorporate experimental data on the charm pair production cross section, only the normalized transverse momentum distribution \( \frac{d\sigma^{\rm norm}}{p_T \, dp_T} \), as shown in Fig.~\ref{lab-fig-dnptdpt}, is taken from FONLL. The complete initial charm-quark transverse-momentum distribution is then written as: \( \frac{d\sigma}{dy \, p_T \, dp_T} = \frac{d\sigma^{\rm norm}}{p_T \, dp_T} \times \frac{d\sigma}{dy},\)
where \( \frac{d\sigma}{dy} \) is the experimentally measured rapidity-differential cross section for charm pairs in $pp$ collisions. This factor does not affect the calculation of \( R_{pPb} \) and \( v_2 \).} Initial charm-quark momenta are randomly sampled based on this distribution. As the medium temperature varies with position, heavy quarks experience different medium effects along their trajectories. Their initial positions are assumed to coincide with regions in the medium where temperature exceeds the critical temperature,  \( T_c = 0.16 \) GeV, representing localized hot spots where charm production is most likely.

\begin{figure}[!hbt]
    \centering
\includegraphics[width=0.4\textwidth]{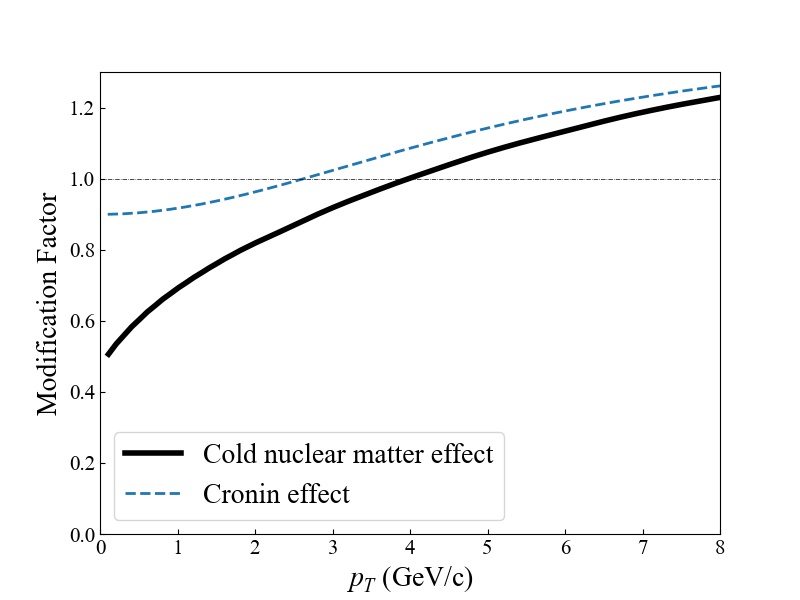}
    \caption{\tred{Modification factors from the Cronin effect and the cold nuclear matter effect, the latter defined as a product of the Cronin factor and the shadowing factor obtained from the EPS09 NLO model. The calculations are performed for $p$-Pb collisions at 8.16 TeV.} }
    \label{lab-cold-fac}
\end{figure}

In \( p \)-Pb collisions, partons undergo multiple scatterings with surrounding partons, gaining additional energy before the hard scatterings in which heavy quark pairs are produced. This results in discrepancies between the initial momentum of heavy quarks in \( p \)-Pb and \( pp \) collisions, a transverse momentum broadening referred to as the Cronin effect. This initial state effect can be incorporated phenomenologically by the following convolution,
\begin{align}
    \overline{f_{pp}(p_T)}={\frac{1}{\pi a_{gN} L}}\int d^2{\bf q_T} e^{-{\frac{q_T^2}{a_{gN}L}}} f_{pp}(|{\bf p_T}-{\bf q_T}|)
\end{align}
where $f_{pp}(p_T)$ is the initial momentum distribution of heavy quarks in \( pp \) collisions. \( L \) is the gluon path length prior to fusion into a heavy quark pair, calculated using the nuclear density modeled by a Woods-Saxon distribution. The Cronin parameter \( a_{gN} \) represents the average transverse momentum squared per unit length of nucleons prior to particle production. A value of approximately \( a_{gN} = 0.15 \ \rm{GeV^2/fm} \) is adopted based on previous studies~\cite{Zhou:2014kka}. \tred{The Cronin modification factor, defined as the ratio of the charm-quark momentum distribution with and without the Cronin effect, is plotted in Fig.~\ref{lab-cold-fac}. For comparision, the total cold nuclear matter modification factor, defined as a product of the Cronin and shadowing factors, is also shown in Fig.~\ref{lab-cold-fac}. 
}

For the shadowing factor, the parton distribution functions in nuclei also differ from the superposition of distributions in free nucleons, leading to the shadowing effect~\cite{Mueller:1985wy}. This alters the parton density, influencing heavy quark and quarkonium production. The modification factor, \( \mathcal{R}_S(x_F, \mu_F) \), is a function of the momentum fraction carried by the gluon in the nucleons \( x_F\), and the factorization scale $\mu_F$~\cite{Zhou:2014kka}, and is therefore dependent on both the rapidity and transverse momentum of the heavy quarks. Shadowing is incorporated using the EPS09 NLO model~\cite{Eskola:2009uj}, applying a rapidity-averaged shadowing factor. Shadowing suppresses the production of heavy quarks at low $p_T$, reducing the $D$ meson nuclear modification factor $R_{pPb}(p_T)$ in this region while minimally influencing azimuthal anisotropy $v_2(p_T)$. 

\tred{Both the Cronin effect and the shadowing effect are classified as initial-state mechanisms that modify the transverse momentum of heavy quarks. The Cronin effect enhanches \( R_{pPb} \) at high \( p_T \), while the energy loss of heavy quarks, categorized as a final-state effect, suppresses \( R_{pPb} \) in the same region within the hot deconfined medium. The combined effects of these contributions leads to an approximately flat \( R_{pPb} \) across \( p_T \). Since neither the Cronin nor shadowing effects alter the initial azimuthal distributions of heavy quarks, variations in the parameters of these initial-state effects have little influence on the \( v_2(p_T) \) of $D$ mesons.}

\section{hydrodynamics}
Extensive studies have established that the hot deconfined matter created in Pb-Pb and \( p \)-Pb collisions is a strongly coupled medium, close to an ideal fluid. Hydrodynamic equations have been widely used to study the expansion of this hot medium. Despite the smaller system size and lower multiplicities in \( p \)-Pb collisions, hydrodynamic models remain applicable in describing the momentum anisotropy of light hadrons.

In this work, the MUSIC package~\cite{Schenke:2010rr}  is used to simulate the evolution of the QCD medium formed in \( p \)-Pb collisions. 
The hydrodynamic equations require closure using the equation of state (EoS) of the medium. For the deconfined phase, the EoS is taken from fit-to-lattice QCD calculations, while the EoS for the hadronic phase is obtained from the Hadron Resonance Gas model~\cite{Huovinen:2009yb}. The two phases are connected by a crossover phase transition. The initial energy density profiles are generated using the Monte Carlo Glauber model~\cite{Miller:2007ri, Shen:2014vra}. Fluctuations in the initial energy density are known to be essential in generating anisotropic flow of light hadrons~\cite{Hirano:2009ah}. To incorporate the influence of these fluctuations on  heavy-flavor evolution, \tred{different anisotropic initial geometries of the initial energy density are parametrized to obtain temperature profiles for use in the Langevin equation.}

\begin{figure}[!hbt]
    \centering
\includegraphics[width=0.47\textwidth]{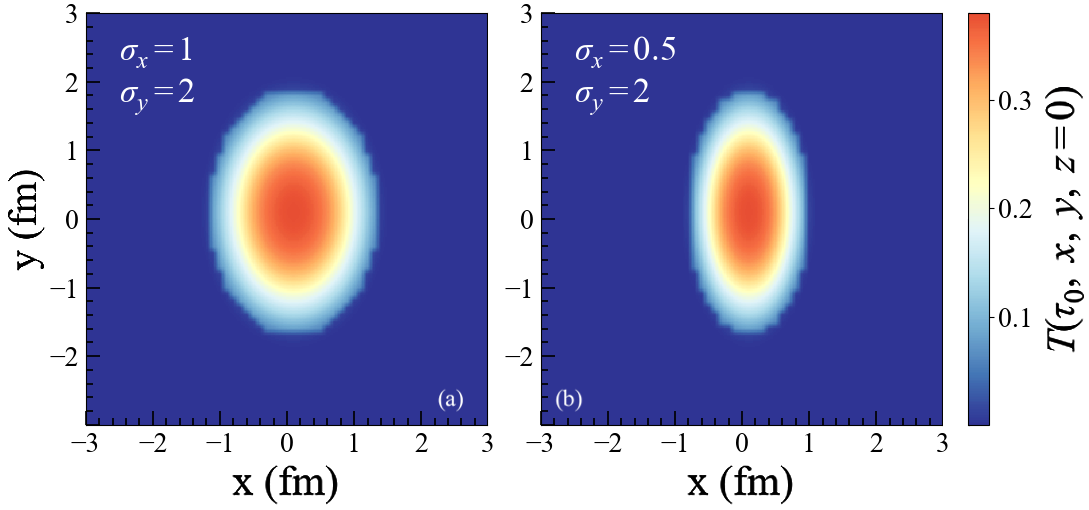}
    \caption{(Color online) Initial temperature profiles in the transverse plane of the medium in 5.02 TeV \( p \)-Pb collisions, with spatial anisotropies $\epsilon_2=0.60$ (a) and $\epsilon_2=0.88$ (b) due to initial fluctuations. }
    \label{lab-smooth-en}
\end{figure}

\tred{The initial energy density of the medium depends on the size of the proton, whose transverse density can be characterized by a Gaussian distribution with a width of around 1 fm, as suggested in Ref. ~\cite{Heinz:2011mh}. Fluctuations distort the shape of the initial energy density, introducing anisotropy.  
To partially incorporate such effects, an elliptic  Gaussian function is used to model the initial energy density $e(x,y, \tau_0)$, }
\begin{align}
    e(x,y,\tau_0) \propto \exp[-\frac{x^2}{\sigma_x^2}-\frac{y^2}{\sigma_y^2}],
    \label{eq-proton-en}
\end{align}
where $\sigma_x$ and $\sigma_y$ determine the eccentricity to which the final momentum anisotropy is proportional. \tred{The Gaussian widths \( \sigma_x \) and \( \sigma_y \), in Eq. (\ref{eq-proton-en}) are chosen to be consistent with proton size.
}Fig.~\ref{lab-smooth-en} shows the initial temperature profiles calculated with varying values of $\sigma_x$ and $\sigma_y$, with eccentricities of $\epsilon_2=0.60$ and $\epsilon_2=0.88$ in the left and right panels, respectively.

\begin{figure}[!hbt]
    \centering
\includegraphics[width=0.48\textwidth]{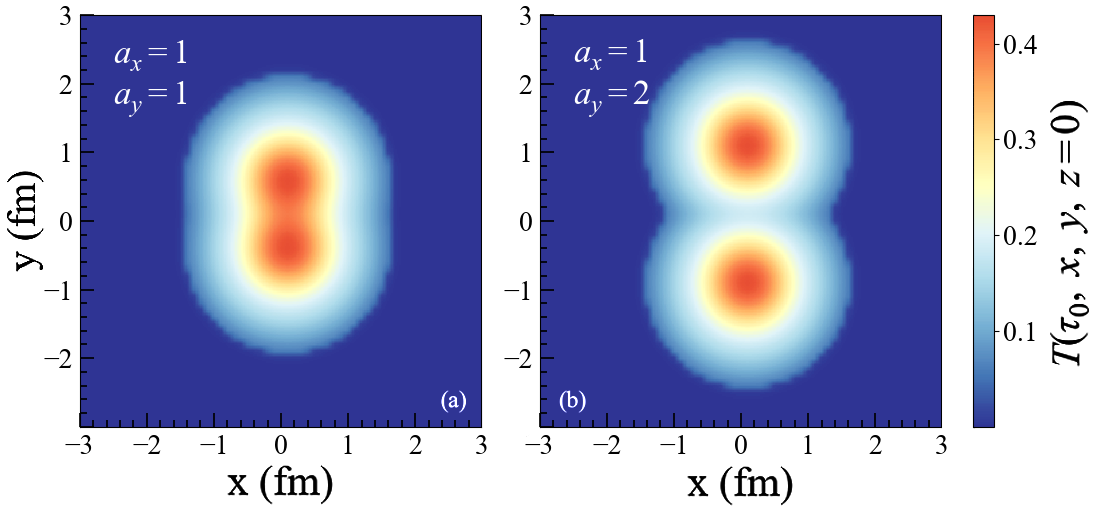}

    \caption{(Color online) Initial transverse-plane temperature profiles of the hot deconfined medium. The Gaussian width of each hot spot is $ a_x = 1$ fm. In panels (a) and (b), the distance between hot spots is set to $a_y=1$ fm ($\epsilon_2=0.23$) and $2$ fm ($\epsilon_2=0.51$), respectively.   }
    \label{lab-DG-fluct}
\end{figure}

\tred{To further investigate the elliptic flow of D mesons in anisotropic small QGP systems, an alternative form of the initial energy density is modeled using multiple Gaussian hotspots:}
\begin{align}
    e(x,y,\tau_0) \propto \sum_{i,j} e^{-{\frac{(x-x_i)^2+(y-y_j)^2}{\sigma^2}}},
\end{align}
with Gaussian width $\sigma=1$ fm for each hot spot. The temperature profiles for configurations of two hot spots with differing separation distances and eccentricities are shown in Figs. \ref{lab-DG-fluct}(a) and \ref{lab-DG-fluct}(b). Their corresponding effects on the nuclear modification factor $R_{pPb}$ and elliptic flow $v_2$ of the $D$ mesons will be presented in Section IV.

\section{numerical results}

In small collision systems, the collective flow of light hadrons is well described by relativistic hydrodynamics, which implies the formation of small QGP droplets. In addition, the azimuthal anisotropy of jets has been studied by considering contributions from quark coalescence and hadronic scattering effects ~\cite{Zhao:2021vmu}. Building upon this foundation, \tred{this work investigates the elliptic flow of D mesons in $p$-Pb collisions by incorporating the anisotropic geometry of the initial energy density, attributed to fluctuations, together with a consistent treatment of the coalescence process in this model.}

\begin{figure}[hbt]
    \centering
\includegraphics[width=0.48\textwidth]{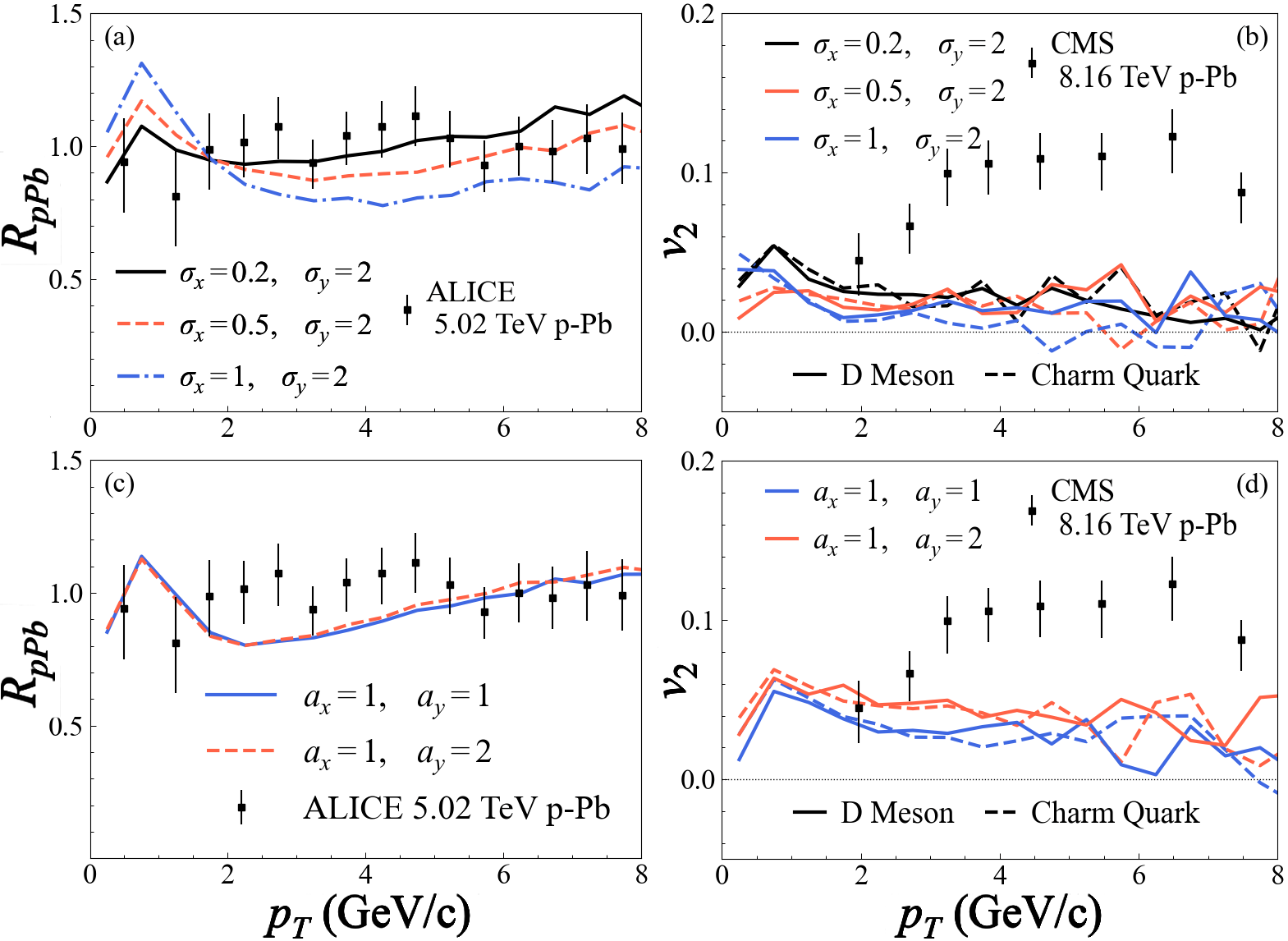}
    \caption{(Color online)  Nuclear modification factor $R_{pPb}$ at \(\sqrt{s_{NN}} = 5.02\) TeV and elliptic flow $v_2$ at \(\sqrt{s_{NN}} = 8.16\) of $D$ mesons in \( p \)-Pb collisions. Results shown in panels (a) and (b) correspond to initial energy densities of the bulk medium proportional to an elliptical Gaussian: \(\exp[-x^2/\sigma_x^2 - y^2/\sigma_y^2]\). Panels (c) and (d) depict results for initial conditions consisting of two Gaussian hot spots with separation distances given by \(a_y\). Experimental data are from~\cite{CMS:2018loe,ALICE:2019fhe}}
    \label{lab-fig-RPA}
\end{figure}

The spatial anisotropy of the initial energy density of the hot medium formed in \( p \)-Pb collisions is characterized by eccentricity $\epsilon_2$. To study its influence on the final momentum of \(D\) mesons in 8.16 TeV $p$-Pb collisions, initial temperature profiles, as shown in Fig.~\ref{lab-DG-fluct}, with varying eccentricities are constructed and coupled to the Langevin equations describing heavy-quark evolution. 
First, an initial energy density described by an anisotropic Gaussian function in the transverse plane of the form \(\exp[-x^2/\sigma_x^2 - y^2/\sigma_y^2]\) is considered (see Fig.~\ref{lab-smooth-en}). A fixed value of \(\sigma_y = 2\) fm and three values for \(\sigma_x = (0.2, 0.5, 1.0)\) fm are used to incorporate initial spatial anisotropies. In Fig.~\ref{lab-fig-RPA}(a), the nuclear modification factor $R_{pPb}$ of \(D\) mesons in 5.02 TeV \( p \)-Pb collisions is computed using these three temperature profiles corresponding to different eccentricities. 

The influence of both hot and cold nuclear matter combine to shape the final \(R_{pPb}(p_T)\) spectrum. While the energy loss process induced by the hot QGP shifts heavy quarks to lower \(p_T\) regions, cold nuclear matter effects, particularly the Cronin effect, transfer additional energy to the heavy quarks prior to evolution. This transverse momentum broadening suppresses \(R_{pPb}\) in the low \(p_T\) region, pushing quarks toward higher \(p_T\). The competing effects of hot and cold nuclear matter result in an overall flat trend in \(R_{pPb}(p_T)\).

Figure \ref{lab-fig-RPA}(b) presents the elliptic flow $v_2$ of $D$ mesons in 8.16 TeV \( p \)-Pb collisions. In this study, heavy-quark momentum distribution is taken to be initially isotropic. A final-state anisotropy then develops due to the path-length-dependent energy loss experienced by quarks traversing varying paths within the anisotropic medium, contributing approximately 3\% to the charm quark \(v_2\) prior to hadronization. 

\tred{In the hadronization process, heavy quarks recombine with thermal light quarks as described by the coalescence model, where the mass of the light quark is taken to be 0.3 GeV. The elliptic flows of light quarks used in this work, which are comparable to the experimental data do not alter the elliptic flows of D mesons evidently via the coalescence process, due to the comparatively small mass of the light quarks.} The path-length-difference effect induced by the anisotropic geometry of the medium accounts for less than half of the experimental \(v_2\) at low \(p_T\), consistent with observations from jet studies~\cite{Betz:2011tu}. This effect also shows weak \(p_T\) dependence, as indicated by the dashed lines in Fig.~\ref{lab-fig-RPA}(b).

\tred{To further investigate the impact of different anisotropic hot media in $p$-Pb collisions on the \( v_2 \) of D mesons, alternative shapes of the initial energy density are also employed to calculate \( R_{pPb} \) and \( v_2 \), as shown in Figs.~\ref{lab-fig-RPA}(c) and \ref{lab-fig-RPA}(d), respectively. This approach allows exploration of how final-state interactions between heavy quarks and an anisotropic medium may together contribute to the \( v_2 \) of D mesons under different conditions.
}
Two cases are shown, corresponding to mediums consisting of two hot spots separated by \(a_y = 1.0\) and \(2.0\) fm. 
\tred{The resulting \( R_{pPb} \) in Fig.~\ref{lab-fig-RPA},  reproduces most of the data points within their experimental uncertainties. However, it remains challenging to theoretically explain the experimental data for \( v_2(p_T) \). Under the assumption that initial state effects only modify the transverse momentum distribution and do not alter the azimuthal distribution of heavy quarks,  calculations with different anisotropic hot media, where heavy-quark momentum anisotropy is generated through final-state interactions alone, cannot adequately explain the large \( v_2 \) values observed experimentally in D mesons, as shown by the solid lines in Figs.~\ref{lab-fig-RPA}(b,d).} 
Future work will extend this study with complete calculations incorporating both heavy- and light-hadron observables, simulated with event-by-event fluctuating hydrodynamics. A more complete treatment of initial-state effects will also be incorporated to explain the experimental data of D mesons.
\vspace{.2cm}
\section{Summary}
\tred{This work has investigated the roles of initial- and final-state interactions in shaping the nuclear modification factor \( R_{pPb} \) and azimuthal anisotropy $v_2$ of \( D \) mesons in \( p \)-Pb collisions. Heavy-quark propagation in the deconfined medium is described using the Langevin equation, while hadronization is implemented through quark coalescence. Initial-state effects, including the Cronin and shadowing effects, alter the initial transverse-momentum distribution of heavy quarks but do not modify their azimuthal distribution in the present framework. Although heavy quarks experience energy loss in the medium, the combined effects of initial cold nuclear matter effects and final-state energy loss in the medium result in a nearly flat \( R_{pPb}(p_T) \) for D mesons, in agreement with the experimental data. }

\tred{For $D$-meson elliptic flow \( v_2(p_T) \), momentum anisotropy is developed through final-state interactions as heavy quarks traverse different trajectories in the anisotropic medium. The initial energy density anisotropy of the medium, arising from fluctuations, is modeled by varying geometries and eccentricities. Theoretical calculations for \( v_2(p_T) \) with different initial geometries of the bulk medium account for approximately one-third of the experimentally measured $v_2$. Varying the final-state interaction parameters alone cannot reproduce the large \( v_2 \) observed in the experimental data. These results suggest that azimuthal anisotropy originating from initial-state effects is necessary to explain the $D$-meson elliptic flow observed in small collision systems.}

\vspace{.5cm}
{\bf Acknowledgement:} The author thanks Professor Baoyi Chen for the invaluable guidance, helpful discussions, and careful proofreading. Additional thanks to Wenhua Fan for assistance with hydrodynamic temperature profile evolution. 

\bibliographystyle{unsrtnat}
\bibliography{paper}% Produces the bibliography via BibTeX.

\end{document}